\documentclass{svjour3}                     
\smartqed  
\usepackage{graphicx}
\newcommand{\bea}{\begin{eqnarray}}
\newcommand{\eea}{\end{eqnarray}}
\newcommand{\nnb}{\nonumber}
\begin{document}

\title{Effective theory with dibaryons and perturbative pions
}
\subtitle{ 
``Perturbative pions around the unitary limit'' 
}


\author{
Shung-Ichi Ando
}


\institute{
Shung-Ichi Ando \at
Department of Physics Education, Daegu University, 
Gyeongsan 712-714, Republic of Korea \\
              Tel.: +82-(0)53-850-6975\\
              Fax: +82-(0)53-850-6979\\ 
\email{sando@daegu.ac.kr}           
}

\date{Received: date / Accepted: date}

\maketitle

\begin{abstract}

We discuss a role of dibaryon fields in an effective field theory 
to study pion cloud effect around the unitary limit of two nucleon 
systems at low energies. 


\keywords{dibaryon field \and perturbative pions \and unitary limit} 
\end{abstract}

\section{Introduction}
\label{intro}

Nonperturbative renormalization for few-body systems is still 
an active issue in studies in effective field theories (EFTs). 
(See recent works, e.g., Refs.~\cite{zme-12,eg-12} and references therein.) 
A difficulty comes in when one solves the nonperturbative 
Lippmann-Schwinger (LS) equation with  
an $NN$ potential constructed in an EFT:
the potential can be singular, solving the LS equation
with the potential generates another set of infinities, and thus 
those infinities are needed to be renormalized.  

A nonperturbative renormalization procedure may be illustrated 
in a following way~\cite{pbc-ap98,yh-prc05}:
A calculated result $a$ is renormalized by a physical quantity 
$a_{phy.}$ as
\bea
a_{phy.} = a(C_n(\Lambda),J_m(\Lambda))\,,
\label{eq;renormalization}
\eea
where $a_{phy.}$ can be effective range parameters or phase shift data
in the $NN$ scattering,
$C_n$ are low energy constants (LECs) of the 
$NN$ potential, 
and $J_m$ are loop functions. 
The loop functions $J_m$ become infinite in general, and thus 
one needs to choose a regularization scheme which introduces a
renormalization scale $\Lambda$ in both $C_n(\Lambda)$ and $J_m(\Lambda)$.
Provided that a regularization method and a value of the scale $\Lambda$ 
are specified, the LECs $C_n(\Lambda)$ are fixed by using a set of
relations in Eq.~(\ref{eq;renormalization}). 

Because expressions of the renormalization relations 
in Eq.~(\ref{eq;renormalization}) are nonlinear 
in terms of $C_n(\Lambda)$ and $J_m(\Lambda)$ 
and quite complicated,
it would be nontrivial that a perturbative series of the 
$NN$ potential obeys naive counting rules 
under the constrains
of Eqs.~(\ref{eq;renormalization}); rather, it might
be more likely that the counting rules or a validity of
the theory significantly depend on detailed ``specifications''
which one employes in his or her calculation, 
such as 
which physical quantities and renormalization scheme 
are used for renormalization,
whether the scale parameter $\Lambda$ is kept at a finite value 
or sent to infinity,
and so on.  
 
Though the situation is complicated, 
to look at the problem from 
another point of view, 
we discuss an unconventional approach, 
introducing dibaryon 
fields~\cite{ah-prc05,st-prc08}. 
Firstly, let us review some aspects of 
$NN$ potentials with different values 
of pion mass.

\section{Some aspects of $NN$ interactions with various values of pion mass}
\label{sec:1}

We briefly mention some aspects of $NN$ potentials 
with four values of the pion mass,
\bea
m_\pi = 0\,, 
\ \ \
140\,,
\ \ \
198\,,
\ \ \ 
354\,\ {\rm MeV}\,.
\label{eq;mpi}
\eea
First two values above are well known; 
pion mass in chiral limit and physical mass, respectively
(we do not discuss the nuclear forces with the physical pion mass here).

The $NN$ interactions in the chiral limit, $m_\pi^\chi=0$, have been studied
by many authors, but it may be worth mentioning one thing: 
a long-range part of the $NN$ interaction might be considered 
being a longer range in the chiral limit because of exchanging 
a massless particle, 
however, the interaction becomes rather 
a short range 
because of a (massless) pseudoscalar particle. 
One can easily see this, e.g., in one-pion-exchange potential 
at leading order in which momenta in two $\pi NN$ couplings
are cancelled with momentum dependence of the potential pion propagator
in the chiral limit, and thus the interaction becomes a point like,
similar to that in the pionless theory. 

The third value of the pion mass, 
$m_\pi^{crit.}=198$~MeV~\cite{bh-prl03,ehmn-epjc06}, 
in Eq.~(\ref{eq;mpi}) is a critical pion mass at the unitary limit
where scattering length of $S$-wave $NN$ scattering becomes infinite
and deuteron binding energy does nil. 
At this limit, three-body nucleon systems exhibit 
a {\it universal} feature, so-called Efimov effect,
where infinitely many bound states with a series of geometric binding 
energies are accumulated at threshold. 

The last value of the pion mass $m_\pi^{latt.}= 354$~MeV is 
the lightest pion mass in a lattice QCD simulation of $NN$ 
scattering~\cite{bbos-prl06}. 
With this lattice pion mass, very small values of 
the scattering lengths are reported; 
$a_0 = 0.63\pm 0.50$~fm for $^1S_0$ channel and 
$a_1 = 0.63 \pm 0.74$~fm for $^3S_1$ channel.
We note that physical values of them are 
$a_0 = -23.7$~fm and $a_1 = 5.42$~fm 
(at $m_\pi^{phys.}$), significantly larger absolute values
than those from the lattice simulation.
Furthermore, they diverge in the unitary limit (at $m_\pi^{crit.}$).

\section{Perturbative pions around the unitary limit}
\label{sec:2}

Now we discuss a perturbative expansion of pion cloud corrections
around the unitary limit for few-body systems, 
instead of that around the chiral limit, 
and a role of the dibaryon fields. 

The unitary limit is unique because 
an underlying theory of the few-body systems 
is totally masked by the universality, 
and thus one can never study QCD in few-nucleon systems in the limit. 
Dynamics of QCD in few-nucleon systems can 
be investigated in a {\it deviation} from it.   
In addition,
Wigner's SU(4) symmetry can be realized in this limit~\cite{w-pr37}. 
Furthermore, the unitary limit for two-body systems is 
the critical point where 
the infinite scale (the infinite $S$-wave scattering lengths) 
appear. In lattice QCD studies of the $NN$ scattering,
one should bring down his or her numerical results 
from the present large pion mass, $m_\pi^{latt.}=354$~MeV, 
to the physical pion mass, $m_\pi^{phys.}=140$~MeV,
crossing over the critical point at the critical 
pion mass, $m_\pi^{crit.}=198$~MeV, where the infinite scale
might appear. 
Therefore, it may be more convenient 
in the few-body studies to choose 
the unitary limit than the chiral limit 
as a theoretical starting point. 

To introduce the unitary limit into a theory treating pions perturbatively, 
we make use of the dibaryon fields (some details can be found 
in Ref.~\cite{ah-prc12}). 
A naive picture of this choice for the $NN$ scattering 
is that in the conventional approach
(its theoretical starting point is in the chiral limit)
we have two nucleon cores, which has a baryon number, $B=1$, and 
infinitely heavy mass, surrounded by
the pion cloud and they interact with each other, 
whereas in the unconventional 
approach with the dibaryon fields (that is in the unitary limit), 
we have one core, which has the baryon number, $B=2$, 
and nil-binding energy or infinite scattering length, 
surrounded by the pion cloud. 
It may be interesting to employ this picture in studying the few-body
systems.

\section{EFT with dibaryons and perturbative pions}
\label{sec:3}

The EFT with dibaryon fields and perturbative pions, 
briefly discusses above, is applied to studying
effective range corrections in the $S$-wave $NN$ scattering 
at low energies~\cite{ah-prc12}.

Though we employ the dibaryon fields to incorporate the picture
discussed above, as discussed in the introduction, 
our result would depend on details of the calculation. 
Specifications of our calculation are in the following:  
We calculate $NN$ scattering amplitudes for $^1S_0$ and $^3S_1$ channels,
treat the pions perturbatively and consider only leading one-pion-exchange 
diagrams around the leading order amplitude generated by the dibaryon fields, 
employ the dimensional regularization (DR) and PDS scheme 
when calculating one-loop diagrams, and renormalize parameters 
by using two effective range parameters, scattering length $a$ and 
effective range $r$, for each of the $S$-wave channels.  

Effective range parameters of $S$ wave $NN$ scattering 
are given as 
\bea
p \cot\delta_0 
= - \frac{1}{a} 
+ \frac12 rp^2
+ v_2 p^4
+ v_3 p^6 
+ v_4 p^8 
+ \cdots\,,
\eea
where $p$ is the relative momentum of two nucleons, $\delta_0$ are 
$S$-wave phase shifts, and $v_2$, $v_3$, and $v_4$ are effective
range parameters in higher orders. 
After the renormalization of $a$ and $r$, we obtain expressions
of the effective range parameters $v_2$, $v_3$, and $v_4$ as
\bea
v_2 &=& \frac{g_A^2m_N}{16\pi f_\pi^2}\left\{
-\frac{16}{3}
\frac{1}{a_d^2(\mu)m_\pi^4}
+ \frac{32}{5}
\frac{1}{a_d(\mu)m_\pi^3}
- \frac{2}{m_\pi^2}\left[1+\frac{r_d(\mu)}{a_d(\mu)}\right]
+ \frac43 \frac{r_d(\mu)}{m_\pi}
\right\}\,,
\\
v_3 &=& \frac{g_A^2m_N}{16\pi f_\pi^2}\left\{
16 \frac{1}{a_d^2(\mu)m_\pi^6}
- \frac{128}{7}
\frac{1}{a_d(\mu)m_\pi^5}
+ \frac{16}{3}\frac{1}{m_\pi^4}
\left[ 1 + \frac{r_d(\mu)}{a_d(\mu)}\right]
\right. \nnb \\ && \left.
- \frac{16}{5}\frac{r_d(\mu)}{m_\pi^3}
+\frac12\frac{r_d^2(\mu)}{m_\pi^2}
\right\}\,,
\\
v_4 &=& \frac{g_A^2m_N}{16\pi f_\pi^2}\left\{
-\frac{256}{5}\frac{1}{a_d^2(\mu)m_\pi^8}
+\frac{512}{9}\frac{1}{a_d(\mu)m_\pi^7}
-16\frac{1}{m_\pi^6}\left[
1 + \frac{r_d(\mu)}{a_d(\mu)}\right]
\right. \nnb \\ && \left.
+\frac{64}{7}\frac{r_d(\mu)}{m_\pi^5}
-\frac43\frac{r_d^2(\mu)}{m_\pi^4}
\right\}\,,
\eea
where $a_d(\mu)$ and $r_d(\mu)$ are {\it bare} scattering length and 
effective range generated from the dibaryon fields, and $\mu$ is a
scale parameter from the DR and PDS scheme.
We obtain new corrections to $v_2$, $v_3$, $v_4$ from the effective 
range $r$ ($r_d(\mu)$ in the above expressions) compared to those
previously obtained by Cohen and Hansen~\cite{ch1-prc99,ch2-prc99}.

We note that though the bare parameters are usually replaced by the 
physical ones, we retain the $\mu$ dependence in them because we 
renormalize the effective range parameters $a$ and $r$ and 
regard the higher order ones $v_2$, $v_3$, and $v_4$ as contributions 
from higher energies depending on the scale parameter $\mu$.
We find a large sensitivity of $v_2$, $v_3$, and $v_4$ to the $\mu$ value.
Subsequently, we fit $v_2$ to a $v_2$ value obtained from partial wave analysis
(PWA) by adjusting the value of $\mu$ where we regard 
the scale $\mu$ as
a fine-tuning parameter of the high energy contributions and have
$\mu = 178$ and 330~MeV for the $^1S_0$ channel and $\mu = 167$ and 246~MeV
for the $^3S_1$ channel.  We find a fairly good agreement of 
the effective range parameters $v_3$ and $v_4$ to the values of PWA 
with those $\mu$ values.  

In our results of the phase shifts $\delta_0$, 
if we require a $\mu$-independence to 
the results of $\delta_0$, 
they agree to those of the accurate potential model calculation
up to $p\sim 50$~MeV mainly because of our chose of renormalization 
method
fixing two effective range parameters $a$ and $r$.
If we use the fine-tuned $\mu$-values mentioned above, the situation is 
improved and the $S$-wave phase shifts show a better agreement with
those from the accurate potential model up to $p\sim m_\pi$.
More details can be found in Ref.~\cite{ah-prc12}.





%
%

\begin{acknowledgements}
The author would like to thank Chang Ho Hyun for collaboration.
This work is supported by the Basic Science Research Program through
the National Research Foundation of Korea (NRF) funded by the 
Ministry of Education, Science, and Technology (2012R1A1A2009430).
\end{acknowledgements}



\end{document}